\documentclass[11pt]{cernrep}
\usepackage{graphicx,epsfig,here}
\def\chioi{\tilde{\chi}^0_1}
\def\chioii{\tilde{\chi}^0_2}
\def\chipm{\tilde{\chi}^{\pm}_1}
\def\staui{\tilde{\tau}_1}
\def\slr{\tilde{l}_R}
\def\sql{\tilde{q}_L}
\def\sqr{\tilde{q}_R}
\def\ETM{E_T^{miss}}

\def\MT2{M_{T2}}

\def\m0{${0}$}

\begin{document}
\title{Measuring the Mass of the Lightest Chargino at the CERN LHC}
\author{M.M. Nojiri$^1$, G. Polesello$^2$ and D.R. Tovey$^3$}
\institute{$^1$ YITP, Kyoto University, Kyoto 606-8502, Japan \\
$^2$ INFN, Sezione di Pavia, Via Bassi 6, 27100 Pavia, Italy \\ $^3$ Department of Physics
and Astronomy, University of Sheffield, Hounsfield Road, Sheffield S3
7RH, UK} 
\maketitle
\begin{abstract}
Results are presented of a feasibility study of techniques for
measuring the mass of the lightest chargino at the CERN LHC. These
results suggest that for one particular mSUGRA model a statistically
significant chargino signal can be identified and the chargino mass
reconstructed with a precision $\sim$ 11\% for $\sim$ 100 fb$^{-1}$ of
data.
\end{abstract}

\section{INTRODUCTION}

Much work has been carried out recently on measurement of the masses
of SUSY particles at the LHC
\cite{atltdr,Hinchliffe:1997iu,Hinchliffe:1999zc,Bachacou:1999zb,
Allanach:2000kt,Abdullin:1998pm}. These measurements can often be
considered to be `model-independent' in the sense that they require
only that a particular SUSY decay chain exists with an observable
branching ratio. A good starting point is often provided by the
observation of an opposite-sign same-flavour (OS-SF) dilepton invariant
mass spectrum end-point whose position measures a combination of the
masses of the $\chioii$, the $\chioi$ and possibly also the
$\tilde{l}^{\pm}$. Observation of end-points and thresholds in
invariant mass combinations of some or all of these leptons with
additional jets then provides additional mass constraints sufficient
to allow the individual sparticle masses to be reconstructed
unambiguously. A question remains however regarding how the mass of a
SUSY particle can be measured if it does not participate in a decay
chain producing an OS-SF dilepton signature. This problem has been
addressed for some sparticles (e.g. for the $\sqr$ \cite{LHCLC}) however
significant exceptions remain. Notable among these is the case of the
lightest chargino $\chipm$, which does not usually participate in
decay chains producing OS-SF dileptons due to its similarity in mass to
the $\chioii$.

In this paper we attempt to measure the mass of the $\chipm$ by
identifying the usual OS-SF dilepton invariant mass end-point arising
from the decay via $\chioii$ of the {\em other} initially produced
SUSY particle (i.e. not the one which decays to produce the
$\chipm$). We then solve the mass constraints for that decay chain to
reconstruct the momentum of the $\chioi$ appearing at the end of the
chain, and use this to constrain the momentum (via $\ETM$) of the
$\chioi$ appearing at the end of the decay chain involving the
$\chipm$. We finally use mass constraints provided by additional jets
generated by this chain to solve for the $\chipm$ mass. The technique
requires that both the decay chain 
$$
\sql \rightarrow \chioii q \rightarrow \slr lq \rightarrow \chioi llq
$$ 
and the decay chain 
$$
\sql \rightarrow \chipm q \rightarrow qW^{\pm} \chioi \rightarrow qq'q''
\chioi
$$ 
are open with significant branching ratios, and that the
masses of the $\chioi$, $\chioii$, $\slr$ and $\sql$ are known. No
other model-dependent assumptions are required however.

\section{SUSY MODEL AND EVENT GENERATION}

The SUSY model point chosen was that used recently by ATLAS for full
simulation studies of SUSY mass reconstruction \cite{fullsim}. This is a
minimal Supergravity (mSUGRA) model with parameters $m_0$ = 100 GeV,
$m_{1/2}$ = 300 GeV, $A_0$ = -300 GeV, $\tan(\beta)$ = 6 and
$\mu>0$. The mass of the lightest chargino is 218 GeV, while those of
the $\sql$, the $\slr$, the $\chioii$ and the $\chioi$ are $\sim$ 630
GeV, 155 GeV, 218 GeV and 118 GeV respectively. One of the
characteristics of this model is that the branching ratio of $\chipm
\rightarrow W^{\pm} \chioi$ is relatively large ($\sim$ 28
\%). Chargino mass reconstruction involving the decay $\chipm
\rightarrow \staui \nu_{\tau}$ (BR $\sim$ 68 \%) is likely to be very
difficult due to the additional degress of freedom provided by the
missing neutrino. Consequently the $W^{\pm}$ decay mode must be used.

The electroweak SUSY parameters were calculated using the ISASUGRA
7.51 RGE code \cite{Baer:1999sp}. SUSY events equivalent to an
integrated luminosity of 100 fb$^{-1}$ were then generated using Herwig
6.4 \cite{Corcella:2000bw,Moretti:2002eu} interfaced to the ATLAS fast
detector simulation ATLFAST 2.21 \cite{atlfast}. With the standard
SUSY selection cuts described below Standard Model backgrounds are
expected to be negligible. An event pre-selection requiring at least
two ATLFAST-identified isolated leptons was applied in order to reduce
the total volume of data.

\section{CHARGINO MASS RECONSTRUCTION}

Events were required to satisfy `standard' SUSY selection criteria
requiring a high multiplicity of high $p_T$ jets, large $\ETM$ and
multiple leptons:

\begin{itemize}

\item at least 4 jets (default ATLFAST definition \cite{atlfast}) with
$p_T$ $>$ 10 GeV, two of which must have $p_T$ $>$ 100 GeV,

\item $\left(\sum_{i=1}^4 p_{T(jet)}^i + \ETM\right)$ $>$ 400 GeV,

\item $\ETM$ $>$ max$\left(100\mathrm{GeV},0.2\left(\sum_{i=1}^4
p_{T(jet)}^i + \ETM\right)\right)$,

\item exactly 2 opposite sign same flavour isolated electrons or muons
with $p_T$ $>$ 10 GeV,

\item no b-jets or $\tau$-jets.

\end{itemize}

Events were further required to contain dileptons with an invariant
mass less than the expected $l^{\pm}l^{\mp}$ end-point position (100.2
GeV) and at least one dilepton + hard jet combination (one for each
combination of the dilepton pair with each of the two hardest jets)
with an invariant mass less than the expected $l^{\pm}l^{\mp}q$
end-point position (501.0 GeV). The smaller dilepton + hard jet
combination then defined which jet (assumed to be from the decay $\sql
\rightarrow \chioii q$) would be used together with the dileptons to
reconstruct the $\chioii$ production and decay chain.

The momentum of the $\chioi$ at the end of the $\chioii$ decay chain
was calculated by solving analytically the kinematic equations
relating the momenta of the decay products (including the $\chioi$) to
the masses of the SUSY particles, which were assumed to be known from
conventional end-point measurements
\cite{atltdr,Hinchliffe:1997iu,Hinchliffe:1999zc,Bachacou:1999zb,
Allanach:2000kt,Abdullin:1998pm}. This process is described in more
detail in Ref.~\cite{mmn} and results in two solutions for the
$\chioi$ momentum for each of the two possible mappings of the
reconstructed leptons to the sparticle decay products. In the present
analysis just one such mapping was assumed with no attempt being made
to select the correct assignment. Two possible solutions for the
$\chioi$ momentum were therefore obtained for each event.

\begin{figure}[thb]
\begin{center}
\epsfig{file=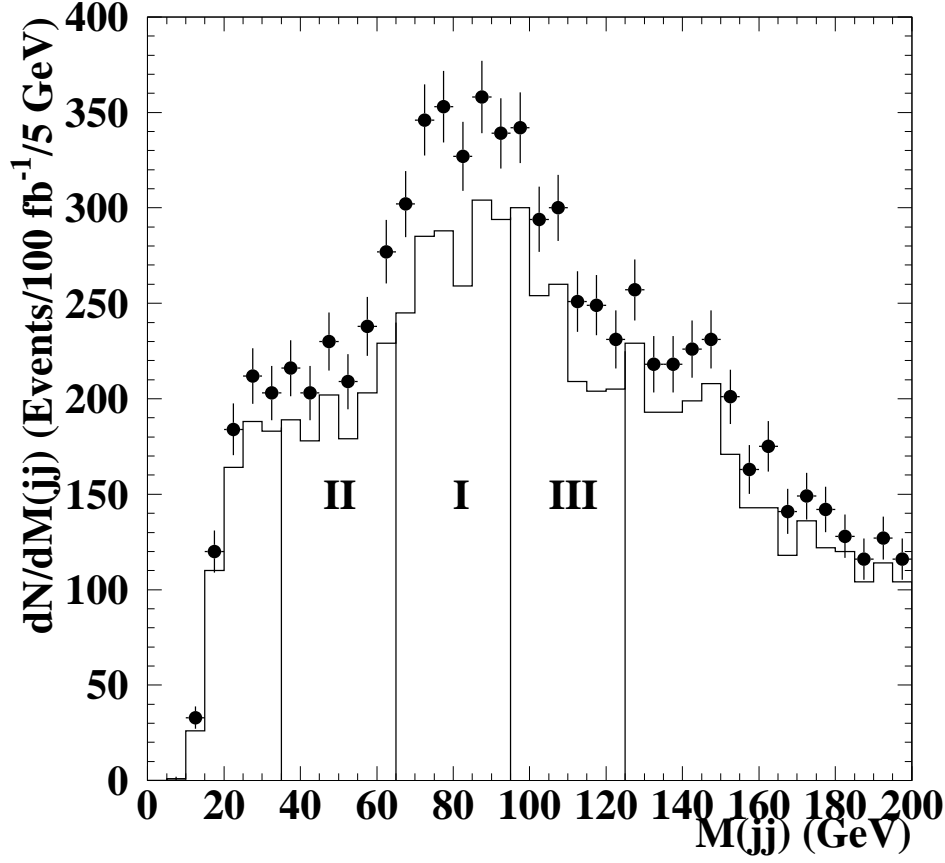,height=5in}
\caption{\label{fig1} {\it Reconstructed dijet invariant mass
distributions for all events (data points) and events not containing
the decay chain $\chipm \rightarrow W^{\pm} \chioi \rightarrow q'q''
\chioi$ selected using Monte Carlo truth. The signal band is labelled
`I' in the figure, while the two sideband are labelled `II' and `III'
respectively.}}
\end{center}
\end{figure}

The nest step in the reconstruction was to find the jet pair resulting
from a hadronic $W^{\pm}$ decay following production via $\chipm
\rightarrow W^{\pm} \chioi$. The potentially large combinatorial
background was reduced by rejecting jet combinations involving either
of the two hardest jets (since these were assumed to arise from $\sql$
decay) and by requiring that the harder(smaller) of the two jets
possessed $p_T$ greater than 40(20) GeV (i.e. selecting asymmetric jet
pairs consistent with a significant boost in the lab frame). A further
cut was applied on the invariant mass of the combination of the jet
pair with the hard jet giving the larger dilepton + jet mass (assumed
therefore to be the jet from the $\sql \rightarrow \chipm q$ decay
preocess). This invariant mass was conservatively required to be less
than that of the $\sql$.

\begin{figure}[thb]
\begin{center}
\epsfig{file=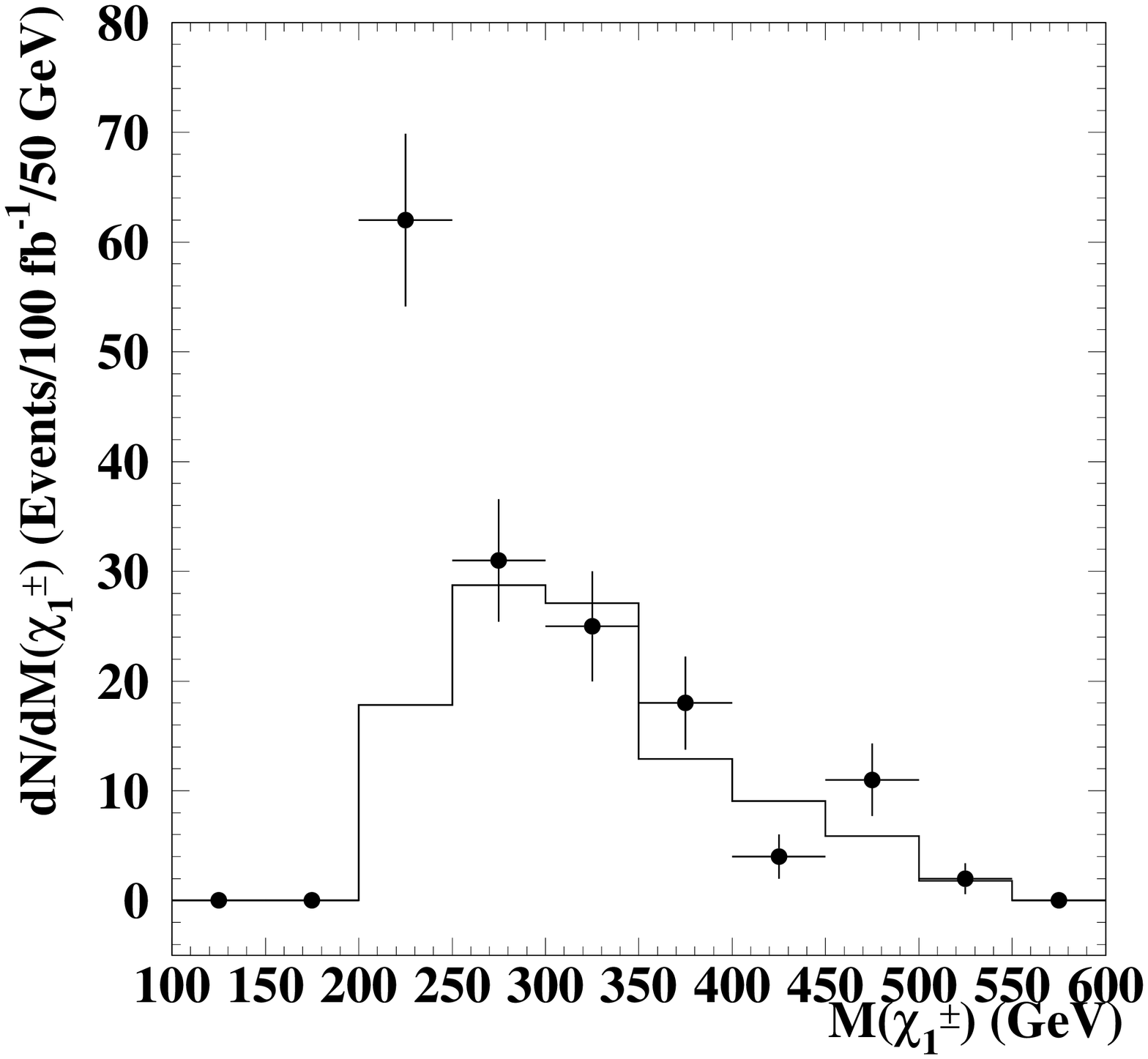,height=2.0in}
\epsfig{file=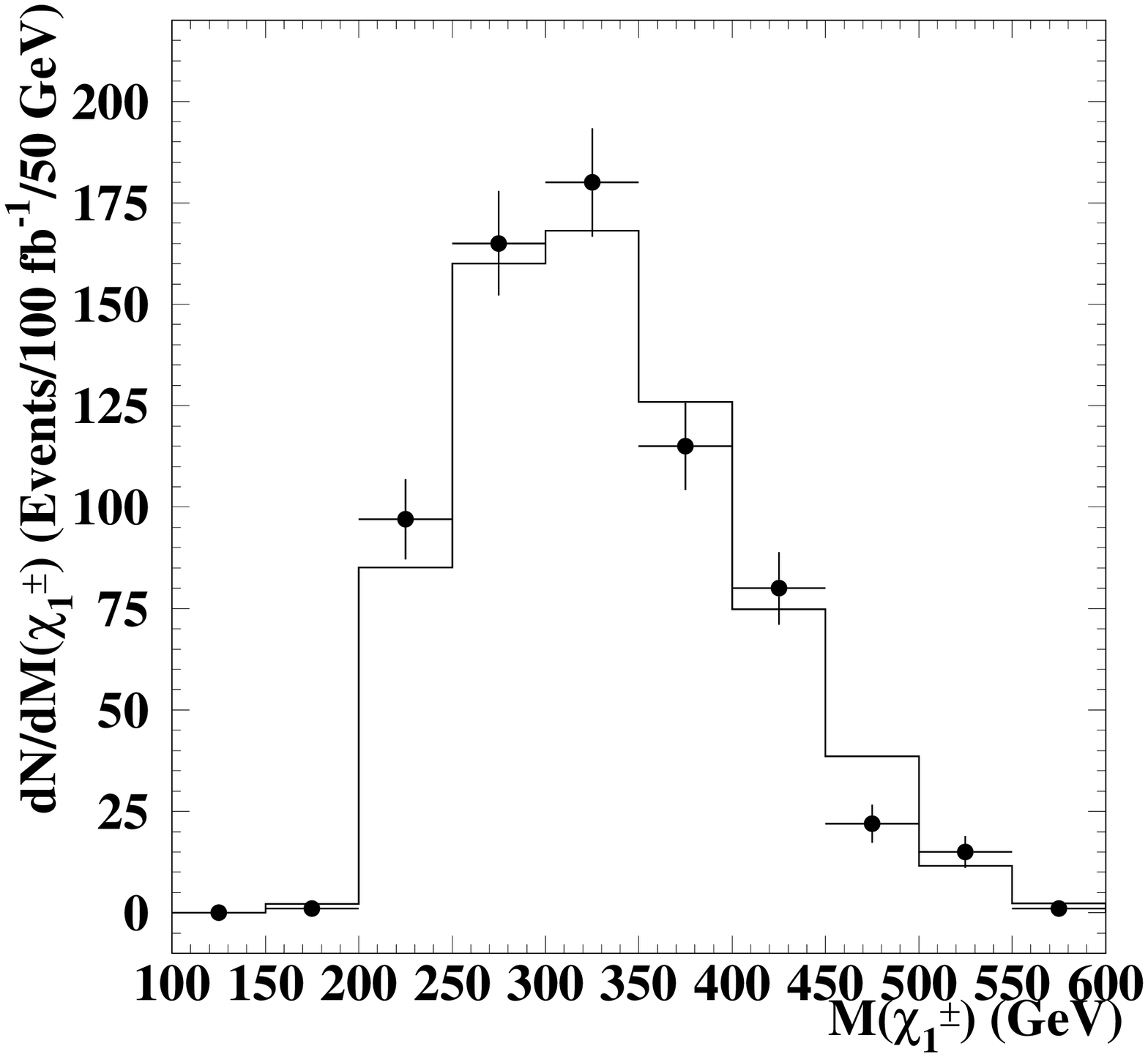,height=2.0in}
\epsfig{file=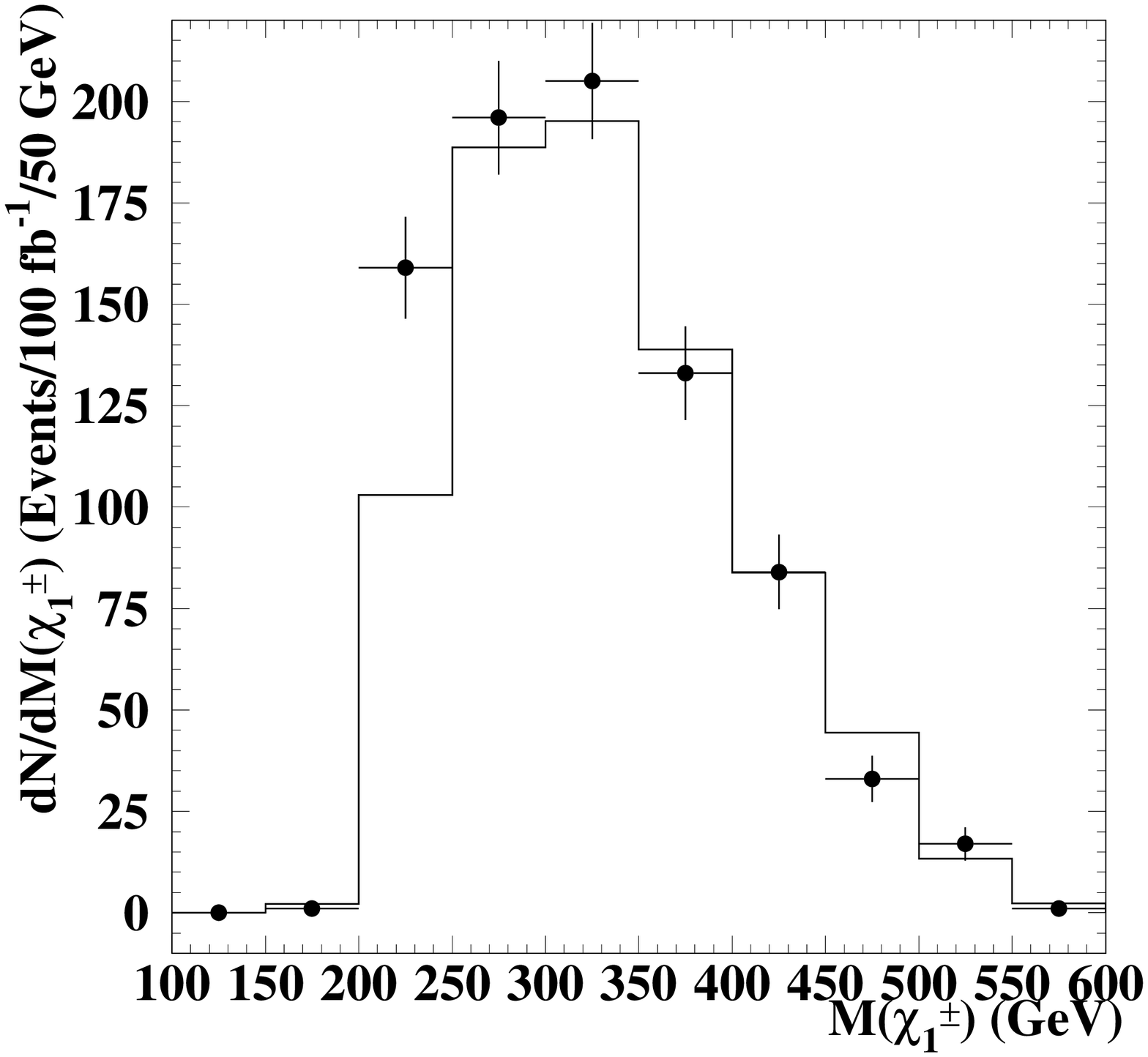,height=2.0in}
\caption{\label{fig2} {\it Reconstructed $\chipm$ mass distributions
showing signal distributions with $|m_{jj}-m_W|$ $<$ 15 GeV (data
points) and sideband distributions with 15 GeV $<$ $|m_{jj}-m_W|$ $<$
45 GeV (histograms). The left hand figure was obtained by selecting
events containing the decay chain $\chipm \rightarrow W^{\pm} \chioi
\rightarrow q'q'' \chioi$ using Monte Carlo truth. The central figure
was obtained by selecting background events not containing this decay
chain. The right hand figure was obtained by using all data.}}
\end{center}
\end{figure}

For each event any jet pairs satisfying the above criteria and
possessing $|m_{jj}-m_W|$ $<$ 15 GeV (Fig.~\ref{fig1}), were
considered to form $W$ candidates. For each event the candidate with
$m_{jj}$ nearest $m_W$ was then selected and used together with the
momentum of the hard jet identified previously and the two assumed $x$
and $y$ components of the $\chioi$ momentum (calculated from the two
solutions for the momentum of the $\chioi$ from the $\chioii$ decay
and $\ETM$) to calculate the chargino mass. Each of the two solutions
for the $\chioi$ momentum gives two possible solutions for
$m_{\chipm}$, the smaller of which is usually physical. Consequently
two possible values for $m_{\chipm}$ were obtained from each event
(plotted in Fig.~\ref{fig2}).

Following this procedure significant backgrounds remain from
combinatorics in SUSY signal events (due to their high average
multiplicity), and from SUSY background events (i.e. events in which
the decay process $\sql \rightarrow \chipm q \rightarrow qW^{\pm}
\chioi \rightarrow qq'q'' \chioi$ is not present). These backgrounds
(or at least those not involving a real $W^{\pm}$ decay) were removed
statistically using a sideband subtraction technique similar to that
described in Ref.~\cite{Hisano:2003qu}. All jet pairs satisfying all
the above selection criteria except the $|m_{jj}-m_W|$ requirement
were recorded if they satisfied the alternative requirement that 15
GeV $<$ $|m_{jj}-m_W|$ $<$ 45 GeV. This requirement then defined two
side-bands located on either side of the main signal band
($|m_{jj}-m_W|$ $<$ 15 GeV) of equal width 30 GeV. The momentum of
each jet pair was then rescaled such that the difference between its
rescaled mass and $m_W$ was the same as the difference between its
original mass and the centre of its sideband (50 or 110 GeV
respectively). Each jet pair was then given a weight of 1.3 (lower
sideband) or 1.0 (upper sideband) to account for the variation of the
background $m_{jj}$ distribution with $m_{jj}$
(Fig.~\ref{fig1}). Values for the chargino mass were then calculated
for each jet pair and used to create a sideband mass distribution
(Fig.~\ref{fig2}). Finally the sideband mass distribution was
subtracted from the signal mass distribution with a relative
normalisation factor of 0.7 to account for the differing efficiencies
for selecting sideband events and background events in the signal
region.

\section{RESULTS}

\begin{figure}[t]
\begin{center}
\epsfig{file=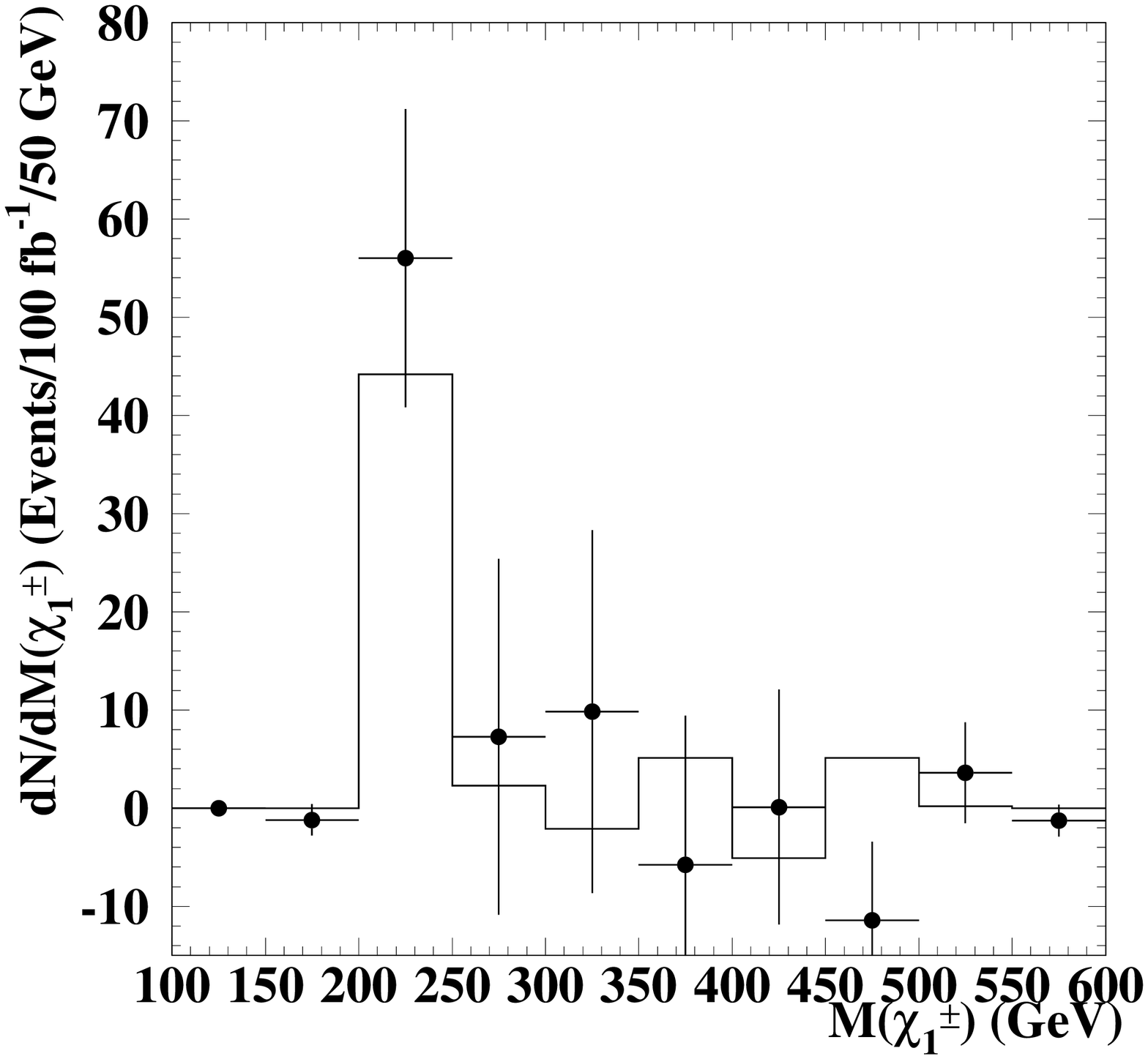,height=5in}
\caption{\label{fig3} {\it Reconstructed $\chipm$ mass distribution
for $\chipm \rightarrow W^{\pm} \chioi \rightarrow q'q'' \chioi$
signal events (histogram) and all events (points with errors).}}
\end{center}
\end{figure}

The sideband subtracted chargino mass distributions obtained from this
process are shown in Fig.~\ref{fig3}, both with and without a
selection requirement for $\chipm \rightarrow W^{\pm} \chioi
\rightarrow q'q'' \chioi$ obtained from Monte Carlo truth. In both
cases no events are observed at masses below the kinematic limit of
198 GeV ($=m_W+m_{\chioi}$) due to the origin of the mass values as
solutions to the kinematic mass relations. In the case where Monte
Carlo truth was used as input a clear peak is seen in the 200 GeV -
250 GeV bin, corresponding well to the actual mass of 218 GeV. At
higher mass values the sideband subtraction process has worked well
and the distribution is consistent with zero. In the case where no
Monte Carlo truth signal event selection has been performed (points
with errors) a clear peak is again seen in the vicinity of the
chargino mass, with few events at higher values. For 100 fb$^{-1}$ the
statistical significance of the peak is around 3 $\sigma$ indicating
that more integrated luminosity (or an improved event selection) would
be required to claim a 5 $\sigma$ discovery. Nevertheless it seems
reasonable to claim that if this data were generated by an LHC
experiment such as ATLAS, and that the observed signal were indeed not
a statistical fluctuation, then the mass of the lightest chargino
could be measured to a statistical precision $\sim$ $\pm$ 25 GeV
($\sim$ 11 \%). More work is needed to determine the likely systematic
error in this quantity arising from effects such as the statistical
and systematic uncertainty in the input sparticle masses used when
calculating the $\chioi$ momentum and $\chipm$ mass.

More work is needed to identify the optimum set of selection criteria
required to identify hadronic $W^{\pm}$ decays in this sample, with
the efficiency of the tau veto (required to remove $\chipm$ decays via
$\staui \nu_{\tau}$) in particular needing to be optimised.  Possible
methods for selecting the correct lepton mapping used to calculate the
$\chioi$ momentum also deserve further study. With these improvements
and/or more integrated luminosity it should then be possible both to
increase the accuracy of the chargino mass measurement and to study
quantities such as the helicity of the $\chipm$ through measurement of
the invariant mass distribution of the $W^{\pm}$ and the hard jet
produced alongside the $\chipm$ in the decay of the parent $\sql$.

\section{CONCLUSIONS}

A study of the identification and measurement of charginos decaying to
$W^{\pm} \chioi$ produced at the LHC has been performed. The results
indicate that for one particular mSUGRA model the mass of the $\chipm$
can be measured with a statistical precision $\sim$ 11 \% for 100
fb$^{-1}$ of integrated luminosity.

\vskip1cm
\noindent

\section*{ACKNOWLEDGEMENTS}
This work was performed in the framework of the workshop: Les Houches
2003: Physics at TeV Scale Colliders. We wish to thank the staff and
organisers for all their hard work before, during and after the
workshop.  We thank members of the ATLAS Collaboration for helpful
discussions. We have made use of ATLAS physics analysis and simulation
tools which are the result of collaboration-wide efforts.  DRT wishes
to acknowledge PPARC and the University of Sheffield for support.

\bibliography{chargino}

\end{document}